# Remembering Professor H. H. Bolotin

**Andrew Stuchbery**
**Professor and Head of Department of Nuclear Physics and Accelerator Applications**
**Research School of Physics, The Australian National University – andrew.stuchbery@anu.edu.au**

*Herb Bolotin was an excellent physicist and educator with a sense of fun and unique humor. He came to Australia as the Chamber of Manufacturer's Professor of Physics at the University of Melbourne in 1971, a position he held until retirement in 1995. He established nuclear structure research using the new 5U Pelletron accelerator in Melbourne, and his group became regular users of the 14UD Pelletron Accelerator at the Australian National University (now part of the Heavy Ion Accelerator Facility). Upon "retirement" in 1996 Herb changed direction and worked successfully on bone densitometry as Emeritus Professor at Melbourne and then as Adjunct Professor at RMIT. He published in this field until 2009. He died suddenly on July 8th 2020.*

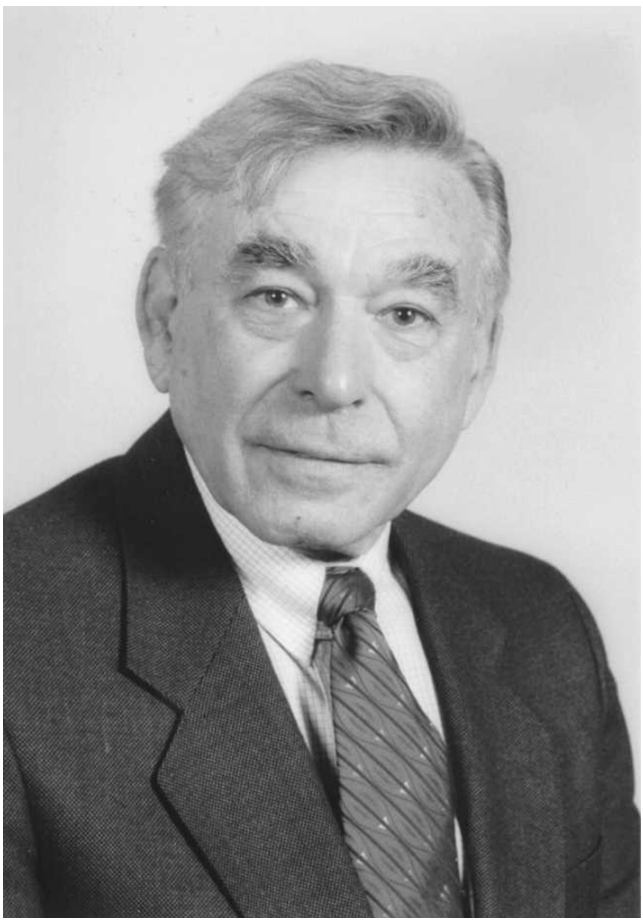

**Figure 1: Professor H. H. Bolotin.**

Herbert Howard Bolotin was born on January 11, 1930 and grew up in New York. He was educated at Stuyvesant High School and the City College of New York (now C.U.N.Y.) – both selective institutions – which specialized in teaching physics, chemistry and mathematics. After he graduated in 1950, he married Charlotte Pearlman in 1951, and was selected as a post-graduate student at Indiana University. He was motivated by the elite scientists there who had worked on the Manhattan Project and were abreast of the newest research and techniques in nuclear physics. He completed his MSc and PhD in 1955.

After graduation he worked for the U.S. Naval Radiological Defense Laboratory (San Francisco); realizing he preferred academia, he held research positions at Brookhaven National Laboratory, Michigan State University, and Argonne National Laboratory. In that period of time, his son Andrew, was born in 1956, and his daughter, Allison in 1960.

After arriving in Australia as Professor of Physics, (Chamber of Manufacturer's Chair), at Melbourne University in 1971, Herb helped to establish the Melbourne Pelletron Laboratory. He pushed to obtain a PDP computer for data acquisition, a progressive step at this time. (A similar computer is shown in the film, "The Dish".) This computer not only supported nuclear structure and nuclear astrophysics research but enabled the development of X-ray imaging with the Melbourne Microprobe. The Pelletron laboratory continues today, featuring microprobe beam lines, and performing materials research. Using this new accelerator, Herb focused initially on measurements of the lifetimes of excited nuclear states using Doppler shift methods. Around 1980, he initiated a program to measure excited-state magnetic moments using the newly developed transient-field technique, and his group became regular users of the 14UD Pelletron Accelerator, now part of the Heavy Ion Accelerator Facility at the Australian National University: aspects of that activity continue today.

Lecturing to undergraduates and honours students, Herb was clear and methodical. His humor, sometimes puzzling to Australian students in the 1970s, punctuated the lectures. For example, a lecture introducing Gauss's Law would invariably contain, "Mr. Gauss says … and by the way, Mr. Gauss is dead." On one occasion Herb asked the class whether the parity of a particular quantum state must be positive or negative. A student took a guess, "Negative?" Herb's response: "Close!"

In a light-hearted moment earlier in his career, Herb had a draftsman put a miniature representation of a lake with a sea serpent into a dip in the drawing of a nuclear decay scheme being submitted for publication. He hoped to sneak it past the referee and editor – but it was spotted! Herb was advised that the paper was accepted for publication subject to the removal of the sea-serpent and lake from the figure. In principle, only the editor and referee should have known about this. But somehow the word got out! For some time afterwards, physicists meeting Herb would ask if he was the "Bolotin of the sea-serpent"? Thus, the joke was shared and enjoyed widely.

Herb had a high regard for the excellent and hard-working graduate students he attracted. With characteristic humor, this regard was usually conveyed by reminding them that they were "slaves" and urging them not to "keep bankers' hours". As a graduate student, writing a paper with Herb was always both an education and a battle over the choice of words and punctuation. The student always lost – except, perhaps, when suggesting the insertion of a semicolon!

Herb was an outstanding mentor, providing opportunities and upholding high standards, always prepared to invest in equipment and infrastructure for research. He sought out and devised clever experimental techniques that helped minimize the possibility for systematic error. Herb had a critical eye to identify where something was going wrong, and a knack for picking topical and important research directions. One example was to measure the lifetimes of excited states in the nucleus platinum-196 [1] just as it was being proposed as the prime example of the novel O (6) limit of the new Interacting Boson Model, which took the nuclear physics world by storm in 1975.

Herb was commissioned by The AGE newspaper in Melbourne, to write a series of articles on any aspect of his field that he found interesting. One of these, entitled "Nuclear Weaponry Self-Taught" was reprinted by the New York Times (December 26 1974). It is available online [2] – a good example of much truth being shared in jest.

Herb's high-quality research using Australian accelerators gained an international profile. As a consequence, he was the first overseas Professor to be invited to do research for 3 months at the new Research Centre for Nuclear Physics at Osaka University, 1977. Later he was invited to Rutgers University (USA), University of California Berkeley (USA), Padua (Italy) and London (UK), where he engaged in research, lectured and mentored students.

In the early 1980s he was invited to give a lecture tour in China by a senior Chinese physicist who felt that his younger colleagues were not keeping up with new ideas and techniques in nuclear physics. Three decades later, while visiting Beijing in 2012, Andrew Stuchbery met Chinese physicists who remembered those lectures.

In 1973, Herb became a member of the Royal Society of Victoria. In 1989, he received the Research Medal from the Royal Society of Victoria; he became a Foundation Fellow in 1995, and later served as its president from 1997-1998. He was elected Fellow of the Australian Institute of Physics (AIP) in 1971. He served as a proactive chair of the Victorian branch of the AIP from 1977 to 1978.

Nearing retirement age, Herb realised he would not be able to mentor a new intake of graduate students at Melbourne University. At this stage, a physician friend drew Herb's attention to the need for a sound floor of physical research under the field of bone densitometry. While winding down his nuclear structure activities, Herb became increasingly interested in this new discipline.

In 1996, as Professor Emeritus, he left Melbourne University, and accepted the post of Adjunct Professor at RMIT Bundoora Campus to work in bone densitometry. The strong leadership of the Chair of Medical Sciences there fostered an atmosphere of collegiality that helped Herb develop his new research focus. Herb's close study of their bone phantoms (models), along with ongoing reading, and speaking to a range of medical friends, soon meant he was able to contribute to this discipline. Over the coming decade he contributed 19 publications including some very well cited papers.

An important example was Herb's seminal paper in the journal *Bone* in 2007 [3], which described significant errors in the DXA (dual-energy x-ray absorptiometry) measurement of bone mineral density with changing body compositions. This work led to more caution in the interpretation of DXA results, helped motivate the development of new technologies in the measurement of bone mineral density, and continues to be cited in 2021.

It is no surprise that Herb's articles on bone densitometry brought the perspective of a rigorous physicist, and focused on experimental methodology, accuracy and scientific procedure.

Herb consistently received ARC grants to support his nuclear structure research, and won NHMRC funding for his work on bone densitometry. Melbourne University awarded him a Doctorate of Science in 1980. During his career he published 100 papers in peer reviewed journals.

Herb died suddenly on July 8th 2020 leaving a solid professional legacy. Recollections of his scientific insight, humour, and ability to get things done will be treasured by his colleagues. His passing leaves an unfillable gap in the hearts of his family and friends who remember him with great affection and respect.

## Acknowledgments

This memoir was prepared with much appreciated assistance from Charlotte Bolotin. The incident concerning the "close" guess of the parity was observed and related by Steve Tims.

## About the author

Andrew Stuchbery is Head of the Department of Nuclear Physics and Accelerator Applications at the Australian National University, which hosts Australia's Heavy Ion Accelerator Facility. Herb Bolotin was his PhD supervisor; he and Herb worked together on nuclear structure physics and related science for almost two decades.

————